# Early $^{56}$Ni decay γ-rays from SN2014J suggest an unusual explosion

Roland Diehl[1]*, Thomas Siegert[1], Wolfgang Hillebrandt[1], Sergei A. Grebenev[3], Jochen Greiner[1], Martin Krause[1], Markus Kromer[4], Keiichi Maeda[5], Friedrich Röpke[6], and Stefan Taubenberger[2]

[1] Max-Planck-Institut für extraterrestrische Physik, Giessenbachstr 1, D-85741 Garching, Germany
[2] Max-Planck-Institut für Astrophysik, Karl-Schwarzschild-Str. 1, 85741 Garching, Germany
[3] Space Research Institute, RAS, Profsoyuznaya 84/32, 117997 Moscow, Russia
[4] The Oskar Klein Centre & Dept. of Astronomy, Stockholm University, AlbaNova, SE-106 91 Stockholm, Sweden
[5] Dept. of Astronomy, Kyoto University, Kitashirakawa-Oiwake-cho, Sakyo-ku, Kyoto 606-8502, and Kavli Institute
       (WPI), University of Tokyo, 5-1-5 Kashiwanoha, Kashiwa, Chiba 277-8583, Japan
[6] Inst. f. Theor. Physik and Astrophysik, Universität Würzburg, Emil-Fischer-Str. 31, 97074 Würzburg, Germany
* Correspondence to: *rod@mpe.mpg.de*            *submitted 11 Apr 2014, revised 10 Jun 2014, accepted 21 Jul 2014*[#] [#]

**Abstract:** Type-Ia supernovae result from binary systems that include a carbon-oxygen white dwarf, and these thermonuclear explosions typically produce 0.5 M$_\odot$ of radioactive $^{56}$Ni. The $^{56}$Ni is commonly believed to be buried deeply in the expanding supernova cloud. Surprisingly, in SN2014J we detected the lines at 158 and 812 keV from $^{56}$Ni decay (τ~8.8 days) earlier than the expected several-week time scale, only ~20 days after the explosion, and with flux levels corresponding to roughly 10% of the total expected amount of $^{56}$Ni. Some mechanism must break the spherical symmetry of the supernova, and at the same time create a major amount of $^{56}$Ni at the outskirts. A plausible explanation is that a belt of helium from the companion star is accreted by the white dwarf, where this material explodes and then triggers the supernova event.

SN2014J was discovered on January 22, 2014 *(1)*, in the nearby starburst galaxy M82, and was classified as a supernova of type Ia (SN Ia) *(2)*. This is the closest SN Ia since the advent of gamma-ray astronomy. It reached its optical brightness maximum on Feb 3, 20 days after the explosion on January 14.75 UT *(3)*. At a distance of 3.5 Mpc *(4)*, a most-detailed comparison of models to observations across a wide range of wavelengths appears feasible, including gamma-ray observations from the $^{56}$Ni decay chain.

Calibrated lightcurves of SNe Ia have become standard tools to determine cosmic distances and the expansion history of the universe *(5)*, but we still do not understand the physics that drives their explosion *(6,7)*. Their extrapolation as distance indicators at high redshifts, where their population has not been empirically studied, can only be trusted if a physical model is established *(5)*. Unlike core-collapse supernovae, which obtain their explosion energy from their gravitational energy, SNe Ia are powered by the release of nuclear binding energy through fusion reactions.

It is generally believed that carbon fusion reactions ignited in the degenerate matter inside a white dwarf star lead to a runaway. This sudden release of a large amount of nuclear energy is enough to overcome the binding energy of such a compact star, and thus causes a supernova explosion of type Ia. A consensus had been for years that the instability of a white dwarf at the Chandrasekhar-mass limit in a binary system with a main sequence or (red-) giant companion star was the most plausible model to achieve the apparent homogeneity *(6)*. However, observations have revealed an unexpected diversity in type-Ia supernovae in recent years *(8)*, and increasing model sophistication along with the re-evaluations of more exotic explosion scenarios have offered plausible alternatives. The consensus now leans towards a broader range of binary systems and more methods of igniting a white dwarf, independent of its mass. De-stabilizing events such as accretion flow instabilities, He detonations, mergers or collisions with a degenerate companion star are being considered *(9-12)*.

Matter consisting of equal numbers of protons and neutrons (such as in carbon and oxygen) binds these nucleons most-tightly in the form of the $^{56}$Ni nucleus. Therefore, $^{56}$Ni is believed to be the main product of nuclear burning under sufficiently hot and dense conditions, such as in SNe Ia. Radioactive decay of $^{56}$Ni then powers supernova light through its gamma-rays and positrons, with a decay chain from $^{56}$Ni (τ~8.8 days) through radioactive $^{56}$Co (τ~111.3 days) to stable $^{56}$Fe. The outer gas absorbs this radioactive energy input and re-radiates it at lower-energy wavebands (UV through IR). But neither the explosion dynamics nor the evolution towards explosions from white dwarf properties and from interactions with their companion stars can easily be assessed, as it remains difficult *(13)* to constrain different explosion models through observations. Some insights towards the nature of the binary companion star have been obtained from pre-explosion data *(8; 14-17)*, and from its interactions with the supernova *(18-20)*.

Gamma-rays can help in constraining inner physical processes. Although initially the SN Ia remains opaque even to penetrating gamma-rays, within several weeks more and more of the $^{56}$Ni decay chain gamma-rays are expected to leak out of the expanding supernova *(21-22)*. The maximum of gamma-ray emission should be reached 70-100 days after the explosion, with intensity declining afterwards due to the radioactive decay of $^{56}$Co *(23)*. Simulations of the explosion and radioactive energy release have been coupled with radiative transport to show that the gamma-ray emission of the supernova is characterized by nuclear transition lines between ~150 and 3000 keV and their secondary, Compton-scattered, continuum from keV energies up to MeV *(22, 24-26)*.

Here we describe the analysis of data from the Spectrometer (SPI) on the INTErnational Gamma-Ray Astrophysics Laboratory (INTEGRAL) space mission with respect to gamma-ray line emission from the $^{56}$Ni decay chain. INTEGRAL *(27)* started observing SN2014J on January 31, 2014 *(28)*, about 16.6 days after the inferred explosion date. We analyzed sets of detector spectra from the Ge detectors of the SPI spectrometer *(29,30)*, collected at the beginning of INTEGRAL's SN2014J campaign. SPI measures photon interaction events for each of its 15 Ge detectors comprising the telescope camera. Imaging information is imprinted through a coded mask selectively shadowing parts of the camera for a celestial source *(29)*. Changing the telescope pointing by ~2.1 degrees after each ~3000 seconds, the mask shadow is varied for the counts contributed from the sky, while the instrumental background counts should not be affected by these small variations in telescope orientation.

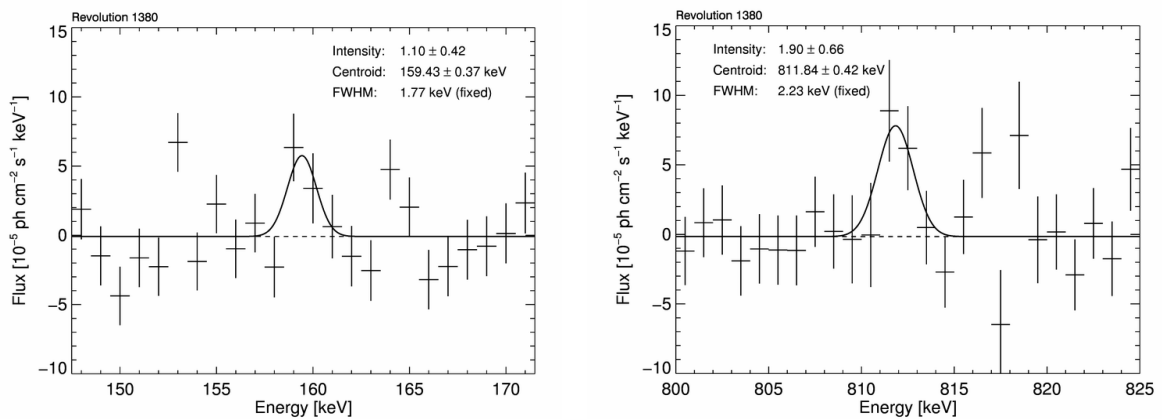

**Fig.1:** Gamma-ray spectra measured with SPI/INTEGRAL from SN2014J. The observed three-day interval around day 17.5 after the explosion shows the two main lines from $^{56}$Ni decay. In deriving these spectra, we adopt the known position of SN2014J, and use the instrumental response and background model. Error bars are shown as 1σ. The measured intensity corresponds to an initially-synthesized $^{56}$Ni mass of 0.06 M$_\odot$.

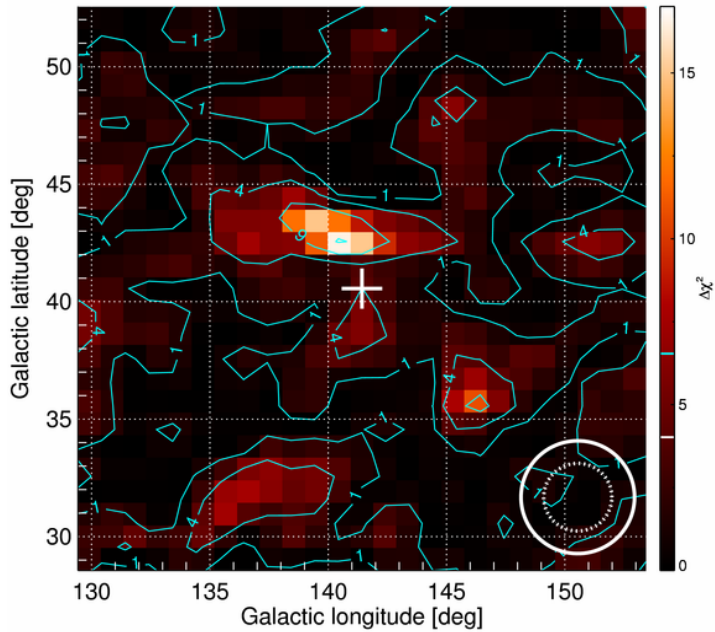

**Fig.2:** Location of the $^{56}$Ni line signal on the sky. The gamma-ray line emission at 158 and 812 keV is mapped onto the position of SN2014J (cross) within instrumental uncertainty (the circle on bottom right shows the size of the instrument point spread function for strong sources). The apparent offset is well within the $2\sigma$ contours of our point spread function, and not significant.

In our analysis of SN2014J, we fitted the measured data to the combination of the expected coding pattern in the detector signals and our instrumental background model. In the fit we adjusted intensities of the global background model and the point source located at a particular position in the sky. The fit wss performed independently for each of the energy bins.

Instrumental background dominates the total count rate in SPI detectors, so its treatment is key to the data analysis. We performed fine spectroscopy of the measured spectra from each 3-day orbit, determining instrumental line features and the status of each detector in terms of intrinsic resolution and degradation. We then used this orbit-averaged spectral shape to fit the line intensity variations among individual pointings, adjusting the underlying continuum emission and the line feature amplitudes relative to the orbit averages. We attributed these line intentisies to each of the Ge detectors as measured in orbit-integrated spectra (for details, see Supplementary Material). In this way, we avoid statistical limitations of invididual-detector spectra per telescope pointing and energy bin, and still obtain a very good representation of background variations, in addition to a consistent description of the spectral evolution of instrumental background with time *(31)*.

The Poissonian maximum likelihood determines the intensity of the celestial source and overall background. For the assumed source position, we derive an intensity spectrum, typically with ~1 keV spectral resolution *(32)*. Our spectra for SN2014J were then analyzed for the presence of lines and their significances. Line parameters such as Doppler shifts or broadening can be derived from fitted line centroids and widths and their uncertainties in a next step. Varying the position of the source in the sky, we also mapped signals in spectral bands of interest across the observed sky area through an identical maximum likelihood analysis. In this way, we can check if a detected line consistently maps to the SN2014J location. From the same sky mapping of a spectral band where we do not identify a celestial line, we obtained a reference that checks for appearance of possible artifacts from statistical fluctuations alone, also accounting for the trials of source positions inherent in such mapping. The dataset analyzed here includes 60 pointings with an exposure of 150.24 ksec from a 3-day period 31 Jan to 2 Feb 2014, i.e., 16.6-19.2 days after the supernova explosion.

We find the characteristic gamma-ray lines of $^{56}$Ni decaying in a SN Ia: the spectra for SN2014J (at Galactic coordinates (l,b)=(141.427°,40.558°)) show the two major lines at 158 and 812 keV (Fig. 1). The two line intensities are identical within uncertainties. Imaging analysis locates the signal at the position of SN2014J, within spatial-resolution uncertainties (Fig. 2). The measured gamma-ray line fluxes are (1.10 ±0.42) $10^{-4}$ ph cm$^{-2}$ s$^{-1}$ (158 keV line) and (1.90 ±0.66) $10^{-4}$ ph cm$^{-2}$ s$^{-1}$ (812 keV line). The observed lines are neither significantly velocity-broadened nor offset (from bulk-velocity Doppler shifts), although broad components could underlie our main signal. Line broadening from the observed $^{56}$Ni corresponds to velocity spreads below ~1500 to 2000 km s$^{-1}$, and its bulk velocity is below ~2000 km s$^{-1}$ (see Supplementary Material, and Table S1.1, for details). The total measured flux can be converted to a $^{56}$Ni mass seen to decay at ~18 days past explosion that corresponds to 0.06 M$_\odot$ of $^{56}$Ni at the time of the explosion.

Apparently, a substantial fraction of the total $^{56}$Ni produced must have been close to the gamma-ray photosphere in the outer ejecta of the supernova at this time, at a depth not exceeding a few g cm$^{-2}$ in column density, and this is probably just the surface of a more-massive concentration of $^{56}$Ni outside the core. Having this much $^{56}$Ni freely exposed is surprising in all explosion models, particularly considering the constraints on its kinematics from line widths and positions.

Observations of SN2014J at early epochs in optical/IR wavelengths also show signatures of an unusual explosion: The rise of supernova light shows that the expected ~t$^2$ behavior *(33)* occurs with some delay. An early steeper rise and possible shoulder may be suggested from observations during hours after the supernova *(20, 3)*, which smooth out after a few days. Also, spectra do not show the early C and O absorption lines that would be expected if a large envelope overlies the light source *(20)*.

A single-degenerate Chandrasekhar-mass scenario appears unlikely from upper limits on X-ray emission that exclude a supersoft progenitor source *(34)*, and from constraints from pre-explosion imaging *(20, 35)*. A sub-Chandrasekhar mass model with a He donor, or a merger of two white dwarfs, may be better models for SN2014J, and could also explain our observation of $^{56}$Ni in the outer layer of the supernova more readily. Moreover, we favor a He donor progenitor channel for this SN Ia, because population-synthesis models (36) indicate a short delay time for such systems, as expected for a SN Ia occurring in a starburst galaxy such as M82. However, a classical double-detonation explosion scenario *(11, 37-38)* is inconsistent with the observations. In this case, a $^{56}$Ni shell engulfing the SN ejecta would be expected, resulting in broad, high-velocity gamma-ray emission lines, whereas narrow lines are detected in SN2014J. However, such an outer shell is expected to have an imprint on optical observables *(39, 33)* that are not seen in SN2014J *(20)*.

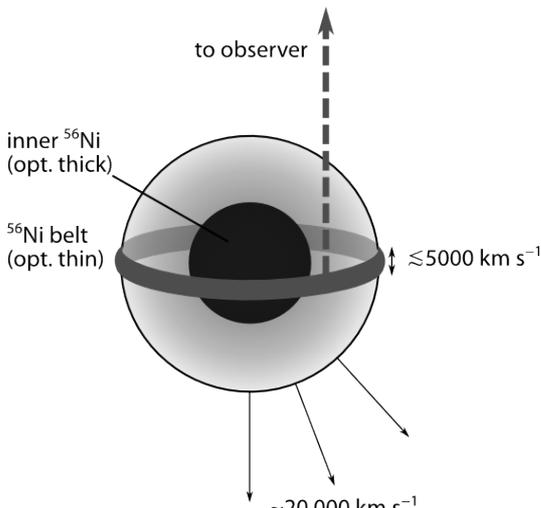

**Fig. 3:** Sketch of a ejecta configuration compatible with our observations. Helium accreted in a belt before the explosion produces the $^{56}$Ni belt at the surface of the ejecta. The gamma-rays can escape from the belt material, while the $^{56}$Ni in the core (black) is still buried at high optical depths. A dashed arrow points to the observer.

A modified but more speculative version of this model may work, however. If He formed an equatorial accretion belt before it detonates, instead of accumulating in a shell, the kinematic constraints could be met, provided we observe the supernova essentially pole-on. In fact, this idea is not new: Accretion belts have been discussed frequently in the context of classical novae *(e.g., 41,42)*. Here we could have the situation of unstable mass transfer on the Kelvin-Helmholtz time scale from the He companion onto the white dwarf, prior to the explosion, which is possible if the He donor is more massive than the white dwarf and fills its Roche lobe. In case of a hydrogen-rich donor, this would be the standard scenario for supersoft X-ray sources *(43)*, which has been excluded for SN2014J *(34)*. But in this case, the accreted material is mostly He, and the accretion rate can be very high, up to 10$^{-4}$ M$_\odot$ y$^{-1}$ *(44)*. If the white dwarf is rapidly rotating or if mass is accreted faster than it loses angular momentum and thus spreads over the white dwarf, a He 'belt' will be accumulated.

An equatorial ring as inferred here might not be that uncommon. Recently, HST imaging of the light echo from the recurrent nova T Pyx revealed a clumpy ring *(45)*. Once this belt becomes dense enough, explosive He burning may be ignited, leaving an ejecta configuration as shown in Fig. 3. This may be consistent with the observed gamma-ray and optical signals. Our radiation transfer simulations in UV/optical/NIR (Fig. S10) shows that the Ni-belt would not produce easily distinguishable features but would result in a normal SN Ia appearance, not only for a pole-on observer but also for an equatorial observer. In view of this, the interpretation of having this type of explosion as a common scenario is not rejected by statistical arguments (see Supplementary Material for more details).

The evolution of the $^{56}$Co gamma-ray signal should reveal further aspects of the $^{56}$Ni distribution in SN2014J. These lines with associated continua have been recognized in data from both INTEGRAL instruments to emerge *(e.g.,46)*, as more of the total $^{56}$Ni produced in the supernova becomes visible when the gamma-ray photosphere recedes into the supernova interior.

## Acknowledgements:

This research was supported by the German DFG cluster of excellence 'Origin and Structure of the Universe', and from DFG Transregio Project No. 33 "Dark Universe". FKR was supported by the DFG (Emmy Noether Programm RO3676/1-1) and the ARCHES prize of the German Ministry for Education and Research (BMBF). The work by K.M. is partly supported by JSPS Grant-in-Aid for Scientific Research (23740141, 26800100). We are grateful to E. Kuulkers for handling the observations, and to X. Zhang for preparing our SPI data, for the INTEGRAL SN2014J campaign. The INTEGRAL/SPI project has been completed under the responsibility and leadership of CNES; we are grateful to ASI, CEA, CNES, DLR, ESA, INTA, NASA and OSTC for support of this ESA space science mission. INTEGRAL's data archive including these data is the ISDC in Versoix, CH, http://www.isdc.unige.ch/integral/archive#DataRelease.